\documentclass[12pt]{amsart}
\usepackage{amsmath, geometry, amssymb} % see geometry.pdf on how to lay out the page. There's lots.
\numberwithin{equation}{section}
\newtheorem{theorem}{Theorem}[section]
\newtheorem{lemma}{Lemma}[section]
\newtheorem{corollary}{Corollary}[section]

\newtheorem{definition}{Definition}[section]
\newtheorem{notation}{Notation}[section]
\newtheorem{example}{Example}[section]

\title{Character Values of the Sidelnikov-Lempel-Cohn-Eastman Sequences}
\author{\c{S}aban Alaca and Goldwyn Millar}
\date{} % delete this line to display the current date

\begin{document}

\maketitle

\begin{abstract} 

Binary sequences with good autocorrelation properties and large linear complexity are useful in stream cipher cryptography. The Sidelnikov-Lempel-Cohn-Eastman (SLCE) sequences have nearly optimal autocorrelation.
However, the problem of determining the linear complexity of the SLCE sequences is still open. 

It is well known that one can gain insight into the linear complexity of a sequence if one can say something about the divisors of the gcd of a certain pair of polynomials associated with the sequence. The authors of \cite{H1}, \cite{K1}, and \cite{M3} were able to obtain some results of this type for the SLCE sequences. The authors of \cite{K1} mention that it would be nice to obtain more such results. We derive new divisibility results for the SLCE sequences in this paper.  

Our approach is to exploit the fact that character values associated with the SLCE sequences can be expressed in terms of a certain type of Jacobi sum. By making use of known evaluations of Gauss and Jacobi sums in the ``pure'' and ``small index'' cases, we are able to obtain new insight into the linear complexity of the SLCE sequences. \\


\end{abstract}

\section{Introduction}

Let $\mathbf{a} = a_0 a_1 a_2 \ldots$ be a sequence over a field $\mathbb{F}$. 
We say that $\mathbf{a}$ is periodic
if there is an integer $v>0$ such that  $a_i = a_{v+i}$ for all integers $i\geq 0$.
If $v$ is the smallest such integer, then we say that $\mathbf{a}$ has period $v$. 
Periodic sequences with certain properties are useful in stream cipher cryptography. A list of general design parameters for cryptographic sequences is given at the end of Section 5.1 in \cite{G1}. A good sequence has a long period and ideally should posses two statistical properties known as the balance property and the run property (Properties R-1 and R-2 from \cite{G1}, respectively). Furthermore, sequences should posses good correlation properties. Individual sequences should have low-valued auto-correlation (Property R-3 from \cite{G1}), 
and sets of sequences should have low-valued cross-correlation. 
Sequences should also have large linear complexity (large linear span). 
We will not discuss the run-property or the low-valued cross-correlation property in this paper.   

It is important that the number of zeroes and ones in the first $v$ elements of a binary sequence of period $v$ differ by at most one \cite{G1}. This is the balance property.

It is possible to define autocorrelation for sequences with elements from various different fields (see \cite{G1}). 
But in this paper, we will discuss only autocorrelation of sequences defined over $\mathbb{F}_2$. 
Thus, we assume that $\mathbf{a}$ is a sequence with elements in $\mathbb{F}_2$. 
We define  the autocorrelation function $C_{\tau}$  of  $\mathbf{a}$ by
$$C_{\tau} = C (\tau ) := \sum_{i = 0}^{v-1} (-1)^{a_i + a_{i + \tau}},$$
where $\tau \in \lbrace 0, ..., v-1 \rbrace$.  
From a cryptographic standpoint, it is important that the maximum autocorrelation of the sequence be as small as possible.

Let $\ell$ be the smallest integer for which there exist $c_1, \ldots ,c_{\ell} \in \mathbb{F}$ such that 
$$-a_i = c_1 a_{i-1} + \cdots + c_{\ell} a_{i - \ell} \mbox{ for each } i \geq \ell.$$ In other words, let $\ell$ be the length of the smallest linear feedback shift register that can be used to generate the sequence $\mathbf{a}$ (see \cite{G1}). 
Then we say that $\ell$ is the linear complexity of $\mathbf{a}.$ 
Linear complexity is one of the most important design parameters for cryptographic sequences: using the Berlekamp-Massey algorithm, one can deduce the entire sequence from $2\ell$ of its consecutive elements \cite{G1}. 
Ideally, the linear complexity of a sequence would be nearly as large as its period. 

The polynomial $c(x) = 1 + c_1x + \cdots  + c_{\ell}x^{\ell} \in \mathbb{F} [x]$ is called the characteristic polynomial 
of  $\mathbf{a}$. 
Let $A(x) = a_0 + a_1x + \cdots + a_{v-1}x^{v-1}.$ It is well known (see for example \cite{G1} and \cite{K1}) that $\mathbf{a}$ has characteristic polynomial 
\begin{eqnarray} \label{c(x)}
c(x) = \frac{x^v-1}{\text{gcd}(x^v -1,A(x))}
\end{eqnarray}
and linear complexity 
\begin{eqnarray} \label{l}
l = v - \text{deg}(\text{gcd}(x^v-1,A(x))).
\end{eqnarray}

As  discussed in \cite{K1}, the computation of $\text{gcd}(x^v-1,A(x))$ is harder when the characteristic of $\mathbb{F}$ divides $v$ than when it does not. For if the characteristic of $\mathbb{F}$ divides $v,$ then one must not only find the common factors of $x^v-1$ and $A(x)$ but also determine the multiplicity with which they divide $\text{gcd}(x^v-1,A(x)).$

In this paper, we  study a class of sequences defined over $\mathbb{F}_2$ that were discovered by Sidelnikov \cite{S1} and rediscovered by Lempel, Cohn, and Eastman \cite{L1}. Following \cite{K1}, 
we refer to these sequences as Sidelnikov-Lempel-Cohn-Eastman sequences (or SLCE sequences). 
As the authors of \cite{K1} remark, SLCE sequences are some of the best even length sequences: they have the same number of zeroes as they do ones, and they have nearly optimal autocorrelation properties \cite{L1}. In fact, since circulant Hadamard matrices seem not to exist \cite{L2}, the autocorrelation properties of the SLCE sequences may in fact be optimal.

We now define the SLCE sequences, and in so doing, we fix notation (for $p,$ $q,$ $m,$ $\alpha,$ $\mathbf{s},$ and $S_2(x)$) 
that we  use throughout the paper. 

\begin{definition} 
Let   $p$  an odd prime,  $m$  a positive integer, and $q = p^m$. 
Let $\alpha$ be a primitive element of the finite field $\mathbb{F}_q.$ 
An SLCE sequence  $\mathbf{s} = s_0s_1 s_2 \ldots$ of period $q-1$ over $\mathbb{F}_2$ is   defined as follows: 

For $0 \leq t \leq q-2,$ we let $s_t := 1$ if  $\alpha^t = \alpha^{2i + 1} - 1$ for some integer $i$ with $0 \leq i \leq q-2$, 
and let $s_t := 0$ otherwise. 
We define $S_2(x) \in \mathbb{F}_2[x]$ by 
$$S_2(x) = s_0 + s_1x + \cdots + s_{q-2}x^{q-2}.$$ 
\end{definition}

Since the SLCE sequences have good autocorrelation and balance properties, it makes sense to study their linear complexity. 
Since these sequences are binary, it is natural to determine their linear complexity over $\mathbb{F}_2$. 
The study of the linear complexity of the SLCE sequences over $\mathbb{F}_2$ began with \cite{H1} and was continued in \cite{K1} and \cite{M3}. However, this problem has turned out to be rather difficult. There are at least two reasons for this. For one thing, since $q-1$ is always even, the characteristic of $\mathbb{F}_2$ divides the periods of the sequences. 
But there is also another problem, which is discussed in the concluding section of \cite{K1}. 
Many well-known sequences correspond (in a sense) to reasonably well-behaved combinatorial objects such as difference sets, divisible difference sets, and partial difference sets (see \cite{B3} for  difference sets and divisible difference sets, and see \cite{M1} for  partial difference sets). 
As a result of this correspondence, explicit formulae have been found for the linear complexity of these sequences (see, for example, \cite{E1}).
However, the SLCE sequences do not correspond to any of these types of combinatorial objects. Rather, they correspond to combinatorial objects called almost difference sets that are, in a sense, more general and about which much less is presently known (see \cite{A3} for background on almost difference sets).

The authors of \cite{H1}, \cite{K1}, and \cite{M3} were able to obtain conditions 
under which certain polynomials divide $\text{gcd}(x^{q-1} + 1,S_2(x)).$ 
In light of (\ref{c(x)}) and (\ref{l}), 
such results provide some insight into the characteristic polynomials of the SLCE sequences over $\mathbb{F}_2$ and yield upper bounds on the linear complexity of these sequences. The authors of \cite{K1} also computed $\text{gcd}(S_2(x),x^{q-1}+1)$ in a number of cases using MAGMA. However, much still remains to be learned about the divisors of these polynomials. Indeed, the authors of \cite{K1} mentioned that it would be nice to obtain new divisibility results giving conditions under which certain polynomials divide $\text{gcd}(S_2(x),x^{q-1}+1).$ We obtain more results of this type in this paper.

The results from \cite{H1} and \cite{K1} are based on a representation of the elements of the SLCE sequences in terms of certain quadratic character values. Using this representation in conjunction with certain facts concerning the cyclotomic numbers of order $2,$ the authors of \cite{H1} and \cite{K1} were able to gain some insight into the characteristic polynomials of these sequences. Furthermore, the authors of \cite{K1} showed that under certain conditions, the problem of determining whether or not a certain polynomial divides $\text{gcd}(x^{q-1}+1,S_2(x))$ is equivalent to determining congruence classes of certain character sums known as Jacobsthal sums. 
The authors of \cite{M3} used known evaluations of cyclotomic numbers in certain special cases to obtain a number of new divisibility conditions.

By contrast, the approach of this paper is based on an expression of character values associated with the SLCE sequences (in a manner to be specified later) in terms of certain Jacobi sums (see Theorem \ref{criterion} below). In fact, the problem of determining whether certain polynomials divide $\text{gcd}(x^{q-1}+1,S_2(x))$ turns out to be equivalent to determining the congruence classes of these Jacobi sums modulo certain prime ideals in certain algebraic number fields.

Jacobi sums are closely related to both cyclotomic numbers and Jacobsthal sums 
(see \cite[Chapters 2 and 6]{B2}), so it is perhaps not surprising that the problem can be interpreted in these various different manners. 
Nonetheless, our method does have some virtues. At present, the Jacobsthal sum condition from \cite{B2} only applies when $q \equiv 1 \pmod{4},$ and calculation of the cyclotomic numbers of order $t$ is quite complicated when $t$ is large. Thus, our representation of the problem in terms of Jacobi sums provides a convenient means by which to harness the information from known evaluations of Gauss and Jacobi sums. Indeed, by making use of such evaluations, we are able to obtain divisibility conditions different than those from \cite{H1}, \cite{K1}, and \cite{M3} 
(see Theorems \ref{nice thm} and  \ref{lengthy conditions} below).

We should also note that since the problem of determining the linear complexity of the SLCE sequences over $\mathbb{F}_2$ is rather difficult, many authors have turned to the important work of calculating the linear complexity of these sequences over other fields. For instance, since the SLCE sequences are constructed using the finite field $\mathbb{F}_q,$ several authors have studied the linear complexity of these sequences over $\mathbb{F}_p$ (see \cite{H2}, \cite{H3}, \cite{G2}, \cite{A2}, \cite{B1}, \cite{K2}, \cite{C1}, \cite{A1}, and \cite{B5}; some of the papers in fact deal with closely related questions). The problem of determining the linear complexity of the SLCE sequences over non-prime fields has also been considered \cite{B4}. 

\section{Preliminary Results}

We introduce some concepts and list some preliminary results that we use throughout the paper.  
Let $G$ denote a finite Abelian group of exponent $v^*$. The integral group ring $\mathbb{Z}[G]$ consists of 
all formal sums $\sum_{g \in G} a_g g$, where $a_g \in \mathbb{Z}$ and with addition and multiplication defined as follows:
\begin{eqnarray*}
\sum_{g \in G} a_g g + \sum_{g \in G} b_g g = \sum_{g \in G} (a_g + b_g) g
\end{eqnarray*}
and
\begin{eqnarray*}
\Big( \sum_{g \in G} a_g g \Big) \Big(  \sum_{h \in G} b_h h \Big) 
= \sum_{f\in G} \Big( \sum_{gh =f} a_g  b_h \Big) f .
\end{eqnarray*}
For any subset $T \subseteq G,$ we identify $T$ with the group ring sum of all the elements in $T$; indeed, we refer to this sum as $T.$

\begin{notation} 
Let $n$ be a positive integer. We  write $\zeta_n$ to denote a primitive, complex $n$th root of unity. 
Sometimes we  write $\zeta$ to refer to a {\em (}not necessarily primitive{\em )} root of unity.
\end{notation} 

A group character is a homomorphism $\chi:G \to \langle \zeta_{v^*} \rangle.$ 
Such a homomorphism can be extended by linearity to a map from $\mathbb{Z}[G]$ to $\mathbb{Z}[\zeta_{v^*}].$ For a discussion of the use of characters in the theory of difference sets, see \cite{B3}; for a discussion of characters over finite fields, see \cite{I1}.

\begin{definition} 
Let $D:= \lbrace \alpha^t \mid  \exists (i \in \{ 0, 1, \ldots, q-2\})~ \alpha^t = \alpha^{2i+1}-1  \rbrace \subseteq \mathbb{F}_{q}^*.$  
We also refer to the group ring element $D \in \mathbb{Z}[\mathbb{F}_q^*]$ as $S_D(\alpha).$
\end{definition}

We adopt the following convention. 
For an integer $i \in \lbrace 1, \ldots ,p-1 \rbrace,$ we  refer to the corresponding element of $\mathbb{F}_p^*$ by italicizing $\mathit{i}.$

\begin{definition} \label{YZ}
 Let $Y:= \lbrace y \in \mathbb{F}_{q}^* \mid  y = x(\mathit{1}-x)   \text{ for some }  x \in \mathbb{F}_{q}^*  \rbrace $. 
Let $Z: = Y^c$ denote the complement of $Y$ in $\mathbb{F}_q^*.$
\end{definition}

The following result, due to Lempel, Cohn, and Eastman \cite[proof of Theorem~5]{L1} plays a fundamental role in  our work.

\begin{theorem} \label{LCE} 
Let $D$ and $Z$ be as in Definitions {\em 2.1} and {\em 2.2}, respectively. Then 
$Z$ is a shift of $D$: in fact, $Z = \mathit{-4}^{-1}D,$ so that $D = \mathit{-4}Z$ and $D^c = \mathit{-4}Y.$
\end{theorem} 

We need several results concerning cyclotomic fields. First, we fix some notation.

\begin{notation} 
Let $k$ be a positive odd divisor of $q-1$, 
and let $f$ denote the multiplicative order of $2$ modulo $k$, 
so that $f$ is the smallest positive integer for which $k|2^f-1$.  
Let $\phi(k)$ denote the Euler phi-function, which is 
the number of positive integers less than $k$ and relatively prime to $k$.
\end{notation}

For a proof of the next result, see \cite[Theorems 8.7 and 8.8]{M2}.

\begin{theorem} \label{FF} 
In the ring of integers $\mathbb{Z}[\zeta_k]$ of the cyclotomic field $\mathbb{Q}(\zeta_k),$ 
the prime ideal factorization of the ideal $\langle 2 \rangle$  is given by 
\[\langle 2 \rangle = P_1P_2 \cdots P_{\phi(k)/f},\] where $P_1, \dots ,P_{\phi(k)/f}$ are distinct prime ideals, 
and for every $i = 1,\ldots ,\phi(k)/f,$ $\mathbb{Z}[\zeta_k]/P_i$ is a finite field of order $2^f.$
\end{theorem}

\begin{notation} 
Let us now stipulate that $\mathcal{P}$ is a prime ideal lying above $2$ in $\mathbb{Z}[\zeta_k].$ 
\end{notation}

For a proof of the following theorem, see \cite[Propositions 13.2.3 and 14.2.1]{I1}.

\begin{theorem} \label{root of unity} The elements $1, \zeta_k, \ldots , \zeta_k^{k-1}$ belong to mutually distinct cosets of $\mathbb{Z}[\zeta_k]/\mathcal{P}.$ Furthermore, if $\gamma \in \mathbb{Z}[\zeta_k]$ and $\gamma \notin \mathcal{P},$ then there exists a unique 
{\em (}not necessarily primitive{\em )} $kth$ root of unity $\zeta$ such that 
$$\gamma^{(2^f-1)/k} \equiv \zeta \pmod{\mathcal{P}}.$$
\end{theorem}

We note that for any quadratic field $K,$ there exists a unique square-free integer $n$ 
such that $K = \mathbb{Q}(\sqrt{n})$, see \cite[p. 95]{A5} or \cite[p. 188]{I1}.
For the proof of the following result, see  \cite[p. 96]{A5} or \cite[p. 189]{I1}. 

\begin{theorem} \label{AI} Let $n \equiv 1 \pmod{4}.$ Let $K = \mathbb{Q}(\sqrt{n})$ be a quadratic field. Then 
the ring $O_K$ of algebraic integers in $K$ is given by
$$O_K = \mathbb{Z} +\mathbb{Z}\big(  \frac{-1+ \sqrt{n}}{2} \big).$$ 
%In other words, $\displaystyle \big\{ 1, \frac{-1+\sqrt{n}}{2} \big\}$  is an integral basis for $\mathbb{Z}[\sqrt{n}].$
\end{theorem} 

%We need the prime ideal factorization of $\langle 2 \rangle$ in certain quadratic fields. 
The following result is a special case of Theorem 10.2.1 from \cite[pp. 242-245]{A5}.

\begin{theorem} \label{prime ideals} Let $K = \mathbb{Q}(\sqrt{n })$ be a quadratic field. 
If $n\equiv 1 \pmod{8},$ then the ideal $\langle 2 \rangle$ factors into a product of two prime ideals as 
\[\langle 2 \rangle = P_1P_2 = \langle 2, \frac{1}{2}(1 + \sqrt{n}) \rangle \langle 2, \frac{1}{2}(1 - \sqrt{n}).\] 
Further, $O_K/P_i$ is a finite field of order $2$  for $i = 1,2$.
\end{theorem}

The following result relates quadratic and cyclotomic fields, see \cite[p. 199]{I1}.

\begin{theorem} \label{quadratic subfield} Let $\ell$ be a prime. 
Then $\mathbb{Q} \big( \sqrt{(-1)^{(\ell - 1)/2} \ell}~ \big)$ is the unique quadratic field contained in the cyclotomic field $\mathbb{Q}(\zeta_{\ell}).$
\end{theorem}

Let $K = \mathbb{Q}(\sqrt{n})$ be a quadratic field. It is known that the set $I(K)$ of all nonzero fractional and integral ideals of $K$ forms an Abelian group under multiplication \cite[Theorem 8.3.4]{A5}. 
Let $P(K)$ be the subgroup consisting of principal ideals. 
The quotient group $H(K) = I(K)/P(K)$ is finite \cite[Theorem 12.5.4]{A5}. 
We call the order of this group the class number of the field $K$ and refer to it as $h(K)$.

We now turn our attention to character sums. 
We note that for every {\rm (}not necessarily primitive{\rm )} $k$th root of unity $\zeta,$ 
there exists a unique character $\chi: \mathbb{F}_q^* \to \langle \zeta_k \rangle$ 
of order dividing $k$ such that $\chi(\alpha) = \zeta$ \cite[Chapter 8]{I1}.

\begin{notation}
Let $\chi: \mathbb{F}_q^* \to \langle \zeta_k \rangle$ denote the unique character mapping $\alpha$ to $\zeta_k$, and 
let $\rho$ be the {\em (}unique{\em )} quadratic character on $\mathbb{F}_q^*.$ 
Note that $\chi$ has order $k$.
\end{notation}

\begin{definition} 
Let $\chi$ be the unique character given above, and let $\phi$ be another nontrivial character of $\mathbb{F}_q^*.$ 
We define the Jacobi sum $J(\chi,\phi)$ by 
\[J(\chi,\phi) := \sum_{i = 1}^{q-2} \chi(\alpha^i)\phi(1 - \alpha^i).\] 
We shall be particularly interested in the Jacobi sum \[K(\chi) := \chi(4)J(\chi,\chi).\]
\end{definition}

We mention the following congruence (see \cite[Theorem 2.18]{B2}).
\begin{equation} 
\label{cong} K(\chi) \equiv -q \pmod{2(1 - \zeta_k)}.
\end{equation}
The following identity is also important for our work  (see \cite[Theorem 2.1.4]{B2}).
\begin{equation} \label{K of chi}
K(\chi) = J(\chi,\rho).
\end{equation}
It is well known that $|J(\chi,\phi)| = \sqrt{q},$ but in general, the exact value of $J(\chi,\phi)$ is not known (and, in particular, the exact value of the Jacobi sum $K(\chi)$ is not known). Such sums have been evaluated in certain special cases. 
For instance, evaluations are known for Jacobi sums over characters of small order. 
This information has already been used to obtain evaluations for cyclotomic numbers \cite[Chapter 2]{B2}
which were in turn used in \cite{M3} to obtain divisibility conditions for the SLCE sequences. 
So, we do not use these evaluations here.

Another case in which evaluations are known is that of the pure Jacobi sums. 
A Jacobi sum is called pure if some positive  integral power of it is real. Such sums were studied in \cite{A4} and \cite{S2}. 
Indeed, in light of  (\ref{K of chi}), 
the results from \cite{A4} and \cite{S2} can be used to evaluate certain Jacobi sums of the type $K(\chi)$.
The authors of \cite{A4} and \cite{S2} showed that if $m$ is odd, then no Jacobi sum defined on $\mathbb{F}_{p^m}$ can be pure. 
They  completely determined conditions under which  Jacobi sums are  pure when $m = 2$. 

\begin{theorem} \label{pure} If $m=2,$ then $K(\chi)$ is pure if and only if $k$ is a divisor of $p+1,$ $k$ is an even divisor of $2(p-1),$ $k=24$ and $p \equiv 17,19 \pmod{24},$ or $k=60$ and $p \equiv 41,49 \pmod{60}.$
\end{theorem}

It follows from Theorem 2.7 and Notation 2.2 that if 
$q=p^2,$ then our sum $K(\chi)$ is pure only when $k$ is an odd divisor of $p+1.$
In this case an explicit evaluation of $K(\chi)$  is given   in \cite[Theorem 2.14]{Br4}.

\begin{theorem} \label{eval} Let $m=2$, and let $k$ be an odd divisor of $p+1.$ Then $K(\chi) = p.$
\end{theorem}

The evaluation in Theorem \ref{eval} is  a special case of a more general result. 
To explain why, it is necessary to introduce another type of character sum.

\begin{definition} Let $\epsilon$ be a character on $\mathbb{F}_q.$ 
We define  the Gauss sum $G(\epsilon)$ by
 \begin{equation} G(\epsilon) :=\sum_{\alpha \in \mathbb{F}_q} \epsilon(\alpha) e^{2\pi i \text{tr}(\alpha)/p},\end{equation} where $\text{tr}$ is the field trace from $\mathbb{F}_q$ to $\mathbb{F}_p.$ 
\end{definition}

The following identity relates Gauss and Jacobi sums (see \cite[Theorem 2.1.3]{I1} or \cite{B2}). 
If $\chi \phi$ is not the trivial character, then 
\begin{equation} \label{Gauss ID} J(\chi,\phi) = \frac{G(\chi) G(\phi)}{G(\chi \phi)}.
\end{equation}
In particular, since $\chi$ is a character of order greater than $2,$ we have 
\begin{equation} K(\chi) = J(\chi,\rho) = \frac{G(\rho)G(\chi)}{G(\chi \rho)}.\end{equation}

Let $s\geq 1$ be an integer,  and let $\chi^{\prime} := \chi \circ N,$ where $N$ is the field norm from $\mathbb{F}_{q^s}^*$ to $\mathbb{F}_q^*$.  Then $\chi^{\prime}$ is a character of $\mathbb{F}_{q^s}$ of order $k$, which is called a lifted character. 
Note that every character on $\mathbb{F}_{q^s}$ of order $k$ can be obtained as a lifted character from a character of $\mathbb{F}_q$ of order $k.$ We mention the following important identity, which is known as the Hasse-Davenport Lifting Theorem (see \cite[Theorem 11.5.2]{B2}).
\begin{equation} \label{Hasse-Davenport} G(\chi^{\prime}) = (-1)^{s-1} (G(\chi))^s.
\end{equation}

The problem of evaluating Gauss sums is just as hard as the problem of evaluating Jacobi sums. But explicit evaluations have been obtained in a number of special cases. The first of these evaluations is due to Gauss, who evaluated $G(\rho)$ when $q = p$ (i.e. when $m=1$). His evaluation can be extended to a general (odd) prime power $q=p^m$ \cite[Theorem 11.5.4]{B2} as 
\begin{equation} \label{Gauss}
 G(\rho) =
  \begin{cases}
   (-1)^{m-1} p^{m/2} & \text{if } p \equiv 1 \pmod{4} \\
   (-1)^{m-1}i^m p^{m/2} & \text{if } p \equiv 3 \pmod{4}.
  \end{cases}
\end{equation} 

A Gauss sum is called pure if some positive integral power of it is real. The following theorem completely classifies pure Gauss sums (see \cite[Section 11.6]{B2}).
\begin{theorem} \label{pure Gauss} Let $n|q-1,$ and let $\epsilon$ be a character of order $n.$ Then $G(\epsilon)$ is pure if and only if there exists a positive integer $x$ such that $p^x \equiv -1 \pmod{n}.$ 
Furthermore, if there exist such integers and $t$ is the least such integer, then there exists 
a positive integer $s$ such that $m = 2ts,$ and 
\[G(\epsilon) = (-1)^{s-1+(p^t+1)s/n} p^{m/2}.\]
\end{theorem}

We now assume that there exists a positive integer $x$ such that $p^x \equiv -1 \pmod{k};$ indeed, we refer to the least such integer as $t.$ 
Then, by Theorem 2.9, $G(\chi)$ is a pure Gauss sum. 
Since $k$ is odd and $p^t + 1$ is even, then $k|p^t+1 \iff 2k|p^t+1.$ 
Hence, since $t$ is the smallest positive integer satisfying $p^t \equiv -1 \pmod{k},$ then $t$ is also the smallest positive integer satisfying 
$p^t \equiv -1 \pmod{2k}.$ 
We note that $\chi \rho$ is a character of order $\text{lcm}(2,k) = 2k.$ 
Thus, $G(\chi \rho)$ is a pure Gauss sum. 
Thus, in this case, 
we can use Theorem \ref{pure Gauss}, (\ref{Gauss}), and (\ref{Gauss ID}) to evaluate the Jacobi sum $K(\chi).$ 
We note that by Theorem \ref{pure Gauss},  $m = 2ts$ for some positive integer  $s $. 
Since the evaluation in (\ref{Gauss}) breaks into two cases, our evaluation also breaks into two cases. 

First, we assume that $p \equiv 1 \pmod{4}.$ Then 
\begin{eqnarray*}
K(\chi) = \frac{(-1)^{m-1} p^{m/2} (-1)^{s-1 + (p^t+1)s/k}p^{m/2}}{(-1)^{s-1 + (p^t+1)s/(2k)}p^{m/2}}  
= (-1)^{1 + (p^t+1)s/(2k)}p^{m/2}.
\end{eqnarray*}
Let us consider the special case in which $m=2$ and $k|p+1$ (so that $t = s = 1$). 
Since $p \equiv 1 \pmod{4}$, it follows that $(p^t+1)/2k$ is odd. 
Then by Theorem \ref{eval}, the evaluation of $K(\chi)$ given above reduces to the evaluation $K(\chi) = p$. 

Next, we assume that $p \equiv 3 \pmod{4}.$ Then 
\begin{eqnarray*}
K(\chi) = \frac{(-1)^{m-1}{i}^{m}p^{m/2}(-1)^{s-1 + (p^t+1)s/k}p^{m/2}}{(-1)^{s-1 + (p^t+1)s/(2k)}p^{m/2}} 
= (-1)^{1 + m/2 + (p^t+1)s/(2k)}p^{m/2}.
\end{eqnarray*}
Again, let us consider the special case in which $m = 2$ and $k|p+1$ (so that $s = t = 1$). 
Since $p \equiv 3 \pmod{4},$ it follows that $(p^t+1)/2k$ is even. 
Then by Theorem \ref{eval}, the evaluation of $K(\chi)$ given above reduces to the evaluation $K(\chi) = p$.
\begin{corollary} \label{Jacobi pure} Assume that there exist positive integers $x$ such that $p^x \equiv -1 \pmod{k},$ and let $t$ be the least such integer. Then there exists $s \in \mathbb{N}$ such that $m = 2ts,$ and 
\begin{eqnarray*}
K(\chi) = \begin{cases}
   (-1)^{1 + (p^t+1)s/(2k)}p^{m/2} & \text{if } p \equiv 1 \pmod{4} \\
    (-1)^{1 + m/2 + (p^t+1)s/(2k)}p^{m/2} & \text{if } p \equiv 3 \pmod{4}.
  \end{cases}
\end{eqnarray*}
\end{corollary}

Finally, a third case in which there are known evaluations for Gauss and Jacobi sums is that of the small index Gauss and Jacobi sums. We will discuss the sums $K(\chi)$ in this context. Recall that $\text{Gal}(\mathbb{Q}(\zeta_k)) \cong \left( \mathbb{Z}/k\mathbb{Z}\right)^*.$ Let $\sigma_p \in \text{Gal}(\mathbb{Q}(\zeta_k))$ be the automorphism mapping $\zeta_k$ to $\zeta_k^p.$ Then, since the Froebenius map is an automorphism of $\mathbb{F}_q$ fixing the elements of $\mathbb{F}_p,$ we have that 
\begin{eqnarray*}
&\sigma_p (K(\chi)) &= \sigma_p(\chi(4))\sum_{i = 1}^{q-2} \sigma_p(\chi(\alpha^i))\sigma_p(\chi(1-\alpha^i)) \\
&&= \chi(4^p)\sum_{i = 1}^{q-2}\chi((\alpha^i)^p)\chi(1^p - (\alpha^i)^p) \\
&&= \chi(4)\sum_{i = 1}^{q-2}\chi(\alpha^i)\chi(1 - \alpha^i) = K(\chi).
\end{eqnarray*}
Thus, $K(\chi)$ is in the fixed field of $\sigma_p,$ and by the Fundamental Theorem of Galois Theory, this field has degree $[(\mathbb{Z}/k\mathbb{Z})^*: \langle p \rangle]$ as an extension of $\mathbb{Q}.$ 
Since we know how to evaluate $K(\chi)$ when there exist positive integers $x$ such that $p^x \equiv -1 \pmod{k},$ we can confine ourselves to the case in which there exist no such integers. Having made this assumption, we see that the quotient group $(\mathbb{Z}/k\mathbb{Z})^*/\langle p \rangle$ must contain the (non-identity) element $-1 + \langle p \rangle$ and so (by Lagrange's Theorem) must have even order. 

The small index assumption is the assumption that $[(\mathbb{Z}/k\mathbb{Z})^*:\langle p \rangle]$ is a small positive integer. 
By making this assumption, we can infer that $K(\chi)$ lies in an algebraic number field of small degree, and can therefore use facts about such number fields to evaluate $K(\chi).$ Explicit evaluations have been obtained for Gauss sums in the index $2$ and index $4$ cases. 
It  is sometimes possible to translate these Gauss sum evaluations into evaluations of $K(\chi)$.

Let us assume that $[(\mathbb{Z}/k\mathbb{Z})^*:\langle p \rangle] = 2.$ 
It is easy to see that $$(\mathbb{Z}/k\mathbb{Z})^* \cong \langle p \rangle \times \langle -1 \rangle.$$ 
Thus, $(\mathbb{Z}/k\mathbb{Z})^*$ contains at most $3$ elements of order $2,$ and it follows easily from the Chinese Remainder Theorem that (since $k$ is odd) either $k = \ell_1^{r_1}$ or $k = \ell_1^{r_1}\ell_2^{r2}$ for some odd primes $\ell_1$ and $\ell_2$,  
and some positive integers $r_1$ and $r_2$.

The following evaluation is due to Langevin \cite{L3}. 
We note that the congruence condition $\ell \equiv 3 \pmod{4}$ is actually forced by the index $2$ assumption, as Langevin demonstrates in his paper. Furthermore, the hypothesis in the evaluation below that $\ell > 3$ is only necessary to obtain a nice expression for the Gauss sum in terms of the class number of a certain quadratic field. We have rephrased Langevin's result in the manner in which it was stated in  \cite{X1}.

\begin{theorem} \label{Langevin} 
Let $k = \ell^{r}$, where  $\ell >3$  is a prime congruent to $3 \pmod{4}$ and $r$ is a positive integer.  
We suppose that $[(\mathbb{Z}/k\mathbb{Z})^*:\langle p \rangle] = 2$ and  
$m = \phi(k)/2$. Then
\[G(\chi) = p^{\frac{1}{2} (m - h)}\left(\frac{a + b\sqrt{-\ell}}{2}\right),\]
where $h = h(\mathbb{Q}(\sqrt{-\ell}))$ is the class number of $\mathbb{Q}(\sqrt{-\ell})$,  and the integers $a$ and $b$ satisfy the three conditions
\begin{eqnarray*}
a,b \not\equiv 0 \pmod{p},~4p^h = a^2 + \ell b^2, \text{ and }a \equiv -2p^{\frac{1}{2} (m + h)} \pmod{\ell}.
\end{eqnarray*}
Furthermore, these conditions are sufficient to determine $a$ completely and to determine $b$ up to sign. 
%It is possible to determine the sign of $b$ using Stickleberger's Congruence.
\end{theorem}

In the above formula, in place of the expression $\left( \frac{a + b\sqrt{-\ell}}{2} \right)$,  
Langevin had originally used the expression $a^{\prime} + b^{\prime}\left(\frac{-1 + \sqrt{-\ell}}{2}\right)$,  
where $a^{\prime}, b^{\prime} \in \mathbb{Z}$. Note that 
\[a^{\prime} + b^{\prime}\left(\frac{-1 + \sqrt{-\ell}}{2}\right) = \frac{\left(2a^{\prime} - b^{\prime}\right) + b^{\prime}\sqrt{-\ell}}{2}.\] 
The integers $a$ and $b$ in the version from \cite{X1} (and from Theorem \ref{Langevin} above) are obtained by setting $a = 2a^{\prime} - b^{\prime}$ and $b = b^{\prime}.$ As a result, we also have the condition (not stated explicitly in our version of Theorem \ref{Langevin}) that $a \equiv b \pmod{2}.$

Note also that $[(\mathbb{Z}/2k\mathbb{Z})^*:\langle p \rangle] =2$. 
Xia and Yang have evaluated index $2$ Gauss sums over characters of order $2 \ell^r$ \cite{X1}.
Their result breaks into two separate cases: one in which $\ell \equiv 3 \pmod{8}$ and one in which $\ell \equiv 7 \pmod{8}.$ 
We only make use of the result for the case in which $\ell \equiv 7 \pmod{8}$.

\begin{theorem} \label{Yang} 
Let $k = \ell^{r}$, where  $\ell >3$  is a prime congruent to $7 \pmod{8}$ and $r$ is a positive integer.  
We supppose that $[(\mathbb{Z}/2k\mathbb{Z})^*:\langle p \rangle]=2$ and  $m = \phi(k)/2$.
Let $\epsilon$ be a character on $\mathbb{F}_q$ of order $2k$. 
Then \[G(\epsilon) = (-1)^{r\frac{p-1}{2}\sqrt{(-1)^{(p-1)/2}}}p^{\frac{m}{2}}.\]
\end{theorem}

Let us  make a slight modification to our earlier hypotheses. 
Assume $s$ is a positive integer, and let $m = \phi(k)s/2.$ So, we are now considering a larger class of prime powers $p^m$.  
Let us set $e = \phi(k)/2,$ so that $m = es.$ Let $\ell \equiv 7 \pmod{8}.$
We consider two cases.

Case $1$: $p \equiv 1 \pmod{4}$. By Theorems \ref{Hasse-Davenport},  \ref{Gauss}, and  \ref{Yang}, we have that 
\[K(\chi) =  \frac{(-1)^{es-1}p^{es/2}(-1)^{s-1}p^{(e-h)s/2}\left(\frac{a + b\sqrt{-\ell}}{2}\right)^s}{(-1)^{s - 1 + r(p-1)s/2 + (p-1)s/4}p^{es/2}}.
\]
Since $e$ is odd and $(p-1)/2$ is even, we deduce that 
\[K(\chi) = (-1)^{s-1-(p-1)s/4}p^{(e-h)s/2}\left(\frac{a + b\sqrt{-\ell}}{2}\right)^s.\]    
 
Case $2$: $p \equiv 3 \pmod{4}$. By Theorems \ref{Hasse-Davenport}, \ref{Gauss}, and  \ref{Yang}, we have that 
\[K(\chi) = \frac{(-1)^{es-1+ es/2}p^{es/2}(-1)^{s-1}p^{(e-h)s/2}\left(\frac{a + b\sqrt{-\ell}}{2}\right)^s}{(-1)^{s - 1 + r(p-1)s/2 + (p-1)s/4}p^{es/2}}.
\]
Since $e$ and $(p-1)/2$ are odd, we deduce that 
\[K(\chi) = (-1)^{s-1-rs + (e + 1)s/2}p^{(e-h)s/2}\left(\frac{a + b\sqrt{-\ell}}{2}\right)^s.\]
We collect these observations for later reference.

\begin{corollary} \label{Jacobi Evaluations}
Let $k = \ell^{r}$, where  $\ell $  is a prime congruent to $7 \pmod{8}$ and $r$ is a positive integer.  
We suppose that $[\mathbb{Z}/k\mathbb{Z}:\langle p \rangle ] = 2$ and $m = \phi(k)s/2$, where $s$ is a positive integer.   

If $p \equiv 1 \pmod{4},$ then 
\[K(\chi) = (-1)^{s-1-(p-1)s/4}p^{(e-h)s/2}\left(\frac{a + b\sqrt{-\ell}}{2}\right)^s.\]   

If $p \equiv 3 \pmod{4},$ then 
\[K(\chi) = (-1)^{s-1-rs + (e + 1)s/2}p^{(e-h)s/2}\left(\frac{a + b\sqrt{-\ell}}{2}\right)^s.\]
\end{corollary}

\section{Character Values}

We show that the problem of finding $\text{gcd}(S_2(x),x^{q-1}+1)$ is equivalent to determining the equivalence class of $K(\chi)$ modulo a certain prime ideal. Several authors have previously made use of complex group characters to determine the linear complexity of various classes of sequences (see, for instance, \cite{M5} and \cite{E1}).

\begin{notation}
Since $\mathbb{F}_{2^f}^*$ is a cyclic group of order $2^f-1$, it has a subgroup of order $k$. 
Hence, the polynomial $x^k+1 = (1+x)(1+x + \cdot \cdot \cdot + x^{k-1})$ splits completely over $\mathbb{F}_{2^f}$.
Let $\beta \in \mathbb{F}_{2^f}$ be an element of order $k,$ so that $\beta$ is a root of $1+x+\cdot \cdot \cdot + x^{k-1}.$ Let $I_{\beta}(x)$ be the minimal polynomial of $\beta$ over $\mathbb{F}_2.$ 
\end{notation}

Note that $I_{\beta}(x)|1+x+ \cdot \cdot \cdot + x^{k-1};$ indeed,  $1+x+\cdot \cdot \cdot + x^{k-1}$ is a product of distinct minimal polynomials of elements of $\mathbb{F}_{2^f}$ of order dividing $k.$ 
Since $k|q-1,$ $\beta$ is a root of $x^{q-1} + 1,$ and so $I_{\beta}(x)$ is a factor of $x^{q-1}+1$ (and, indeed, $1+x+\cdot \cdot \cdot + x^{k-1}|x^{q-1}+1$). 
We want to determine whether or not $I_{\beta}(x)$ and/or $1 + x + \cdot \cdot \cdot + x^{k-1}$ divide $S_2(x).$ 
Note that $I_{\beta}(x)|S_2(x) $ if and only if  $S_2(\beta) = 0$,  where $S_2(\beta)$ is an element of $\mathbb{F}_{2^f}$.

By Theorem \ref{FF} we have  $\mathbb{F}_{2^f} \simeq \mathbb{Z}[\zeta_k] / \mathcal{P}.$  
Let $\phi:\mathbb{F}_{2^f} \to \mathbb{Z}[\zeta_k] / \mathcal{P}$ be an isomorphism. 
Of course, $\phi (0) = 0 + \mathcal{P}$ and $\phi (1) = 1 + \mathcal{P}$. 
Since $\beta$ has order $k,$ there exists $\eta \in \mathbb{F}_{2^f}$ such that $\beta = \eta^{(2^f-1)/k},$ so that $\phi(\beta) = \phi(\eta)^{(2^f-1)/k}.$ Consequently, by Theorem \ref{root of unity}, there exists a unique (in this case, primitive) $k$th root of unity congruent to $\phi(\beta) \pmod{\mathcal{P}}.$ 

\begin{notation} Let $\zeta$ denote the unique primitive $k$th root of unity congruent to $\phi(\beta) \pmod{\mathcal{P}}.$ 
Let $\chi$ denote the unique group character mapping $\alpha$ to $\zeta.$ Let $S_z(x)$ be the polynomial in $\mathbb{Z}[x]$ obtained  by replacing each coefficient of $S_2(x)$ with its counterpart \em{(}$0$ or $1${\rm )} from $\mathbb{Z}.$
\end{notation}

We note that $\phi(S_2(\beta))$ is the equivalence class modulo $\mathcal{P}$ containing $S_z(\zeta)$, and  
\[\chi(D) + \mathcal{P} = \chi(S_G(\alpha)) + \mathcal{P} = S_z(\zeta) + \mathcal{P} = \phi(S_2(\beta)).\] Hence, \[I_{\beta}(x)|S_2(x) \iff \chi(D) \equiv 0 \pmod{\mathcal{P}}.\]
Since $\chi$ is nontrivial, we have $\chi(D) = \chi(G - D^c) = -\chi(D^c),$ so that \[\chi(D) \equiv 0 \pmod{\mathcal{P}} \iff \chi(D^c) \equiv 0 \pmod{\mathcal{P}}.\] Hence, 
\begin{equation} \label{modified condition} I_{\beta}(x)|S_2(x) \iff \chi(D^c) \equiv 0 \pmod{\mathcal{P}}.
\end{equation}
Thus, it suffices to consider $\chi(D^c)$ instead of $\chi(D).$

We now prove the result mentioned at the beginning of this section. As we show in the next section, this result enables us to derive several new divisibility results for the SLCE sequences. We proceed by obtaining an expression for $\chi(D^c)$ in terms of $K(\chi).$ 

\begin{theorem} \label{criterion} We have
$$I_{\beta}(x)|S_2(x) \iff \frac{1}{2}(K(\chi) + 1) \equiv 0 \pmod{\mathcal{P}}.$$
\end{theorem}

\proof 
The reasoning in the next two sentences is taken from \cite[Theorem 2.14]{B2},   
where it serves a different purpose. Let $\gamma \in \mathbb{F}_q^*$ be fixed. 
An element $x \in \mathbb{F}_q^*$  satisfies the equation  $x(\mathit{1}-x) = \gamma$ if and only if it satisfies the equation $(\mathit{2}x-\mathit{1})^2 = \mathit{1}-\mathit{4}\gamma.$ 
Hence, the number of solutions of the equation $x(\mathit{1}-x) = \gamma$ in $F_q^*$  is $1 + \rho(\mathit{1}-\mathit{4}\gamma)$, where  $\rho$ denotes the (unique) quadratic character on $\mathbb{F}_q$. 
It follows that every element of $\mathbb{F}_q^*$ is represented either twice or zero times in the form $x(\mathit{1}-x),$ save for $\mathit{4}^{-1},$ which is represented once. This makes sense since there are $q-2$ choices of $x$ for which $x(\mathit{1}-x) \in \mathbb{F}_q^*,$ and $q-2$ is an odd number.
Making use of Theorem \ref{LCE}, we see that   
\begin{eqnarray*}
&\chi(\mathit{-1})K(\chi) &= \chi(\mathit{-4})J(\chi,\chi) = \chi(\mathit{-4})\sum_{x \in \mathbb{F}_q^*} \chi(x)\chi(\mathit{1}-x) \\
&&= \chi(\mathit{-4})\sum_{x \in \mathbb{F}_q^*} \chi(x(\mathit{1}-x)) = \chi(\mathit{-4})\chi \Big( \sum_{x\in \mathbb{F}_q^*} x(\mathit{1}-x) \Big) \\
&& = \chi(\mathit{-4})\chi( 2Y - \mathit{4}^{-1}) = \chi(2D^c - (\mathit{-1})) = 2\chi(D^c) - \chi(\mathit{-1}).
\end{eqnarray*}
So, we deduce that 
\[\chi(D^c) = \frac{1}{2} \chi(\mathit{-1})(K(\chi) + 1).\]
Note that, by (\ref{cong}), $K(\chi) \equiv 1 \pmod{2},$ so that the value we have ascribed to $\chi(D^c)$ is indeed an element of $\mathbb{Z}\left[\zeta_k \right].$ 
The result now follows by equivalence (\ref{modified condition}).$ \qed$

\section{Divisibility Results}

We use Theorem \ref{criterion}, in conjunction with the evaluations of the sums $K(\chi)$ given in Section $2,$ to obtain new results concerning the divisors of $\text{gcd}(S_2(x),x^{q-1}+1).$ 
We first apply the evaluations of the pure Jacobi sums given in Corollary \ref{Jacobi pure}.

\begin{lemma} \label{pure conditions} Suppose that there exist positive integers $x$ satisfying the congruence 
$p^x \equiv -1 \pmod{k},$ and let $t$ be the least such integer. 
Hence, by Theorem {\rm \ref{pure Gauss}}, $m = 2ts$ for some positive integer $s$. 

If $p \equiv 1 \pmod{4},$ then $I_{\beta}(x)|S_2(x) \iff s \equiv 0 \pmod{2}.$ 

If $p \equiv 3 \pmod{4},$ then $I_{\beta}(x)|S_2(x) \iff$ either $s \equiv 0 \pmod{2}$ or $ts$ is odd.
\end{lemma}

\proof
By Corollary \ref{Jacobi pure}, $K(\chi)$ is pure; in fact, $K(\chi) \in \mathbb{Z}.$ 
We know that $\mathcal{P}\cap \mathbb{Z} = 2\mathbb{Z}$ (see \cite{I1}). 
Hence, \[I_{\beta}(x)|S_2(x) \iff \frac{1}{2}(K(\chi)+1) \equiv 0 \pmod{2} \iff K(\chi) + 1 \equiv 0 \pmod{4}.\] 

If $p \equiv 1 \pmod{4},$ then by Corollary \ref{Jacobi pure}, we have 
\begin{eqnarray*}
&I_{\beta}(x)|S_2(x)& \iff (-1)^{1 + (p^t + 1)s/(2k)}p^{m/2} + 1 \equiv 0 \pmod{4} \\
&& \iff (-1)^{1 + (p^t+1)s/(2k)} + 1 \equiv 0 \pmod{4}.
\end{eqnarray*}
Since $k$ is odd, we have
\[I_{\beta}(x)|S_2(x) \iff (-1)^{1 + s} + 1 \equiv 0 \pmod{4} \iff s \equiv 0 \pmod{2}.\]

If $p \equiv 3 \pmod{4},$ then by Corollary \ref{Jacobi pure}, we have 
\[I_{\beta}(x)|S_2(x) \iff (-1)^{1 + m/2 + (p^t + 1)s/(2k)}p^{m/2} + 1 \equiv 0 \pmod{4}.\] 
We first assume that $ts$ is even. Thus, $p^{m/2} \equiv 1 \pmod{4}.$ 
Hence, \[I_{\beta}(x)|S_2(x) \iff (-1)^{1 + (p^t + 1)s/(2k)} + 1 \equiv 0 \pmod{4}.\] 
If $t$ is even and $s$ is odd, then $1 + (p^t + 1)s/(2k) \equiv 0 \pmod{2}.$ 
On the other hand, if $s$ is even, then $1 + (p^t + 1)s/(2k) \equiv 1 \pmod{2}.$ 
Hence, if $ts$ is even, then 
\[I_{\beta}(x)|S_2(x) \iff s \equiv 0 \pmod{2}.\]
We now assume that $ts$ is odd. Then \[I_{\beta}(x)|S_2(x) \iff (-1)^{ts + (p^t+1)s/(2k)} + 1 \equiv 0 \pmod{4} \iff (-1) + 1 \equiv 0 \pmod{4}.\] So, clearly $I_{\beta}(x)|S_2(x)$ when $ts$ is odd.$\qed$

We use Lemma \ref{pure conditions} to determine conditions under which $1 + x + \cdot \cdot \cdot + x^{k-1} \mid S_2(x).$

\begin{theorem} \label{nice thm} Suppose that there exist positive integers $x$ satisfying the congruence $p^x \equiv -1 \pmod{k},$ 
and let $t$ be the least such integer. 
Hence, by Theorem {\rm \ref{pure Gauss}}, $m = 2ts$ for some positive integer $s$. 

If $p \equiv 1 \pmod{4},$ then $1 + x + \cdots + x^{k-1}|S_2(x) \iff s \equiv 0 \pmod{2}.$ 

If $p \equiv 3 \pmod{4},$ then $1 + x + \cdots + x^{k-1}|S_2(x) \iff$ either $s \equiv 0 \pmod{2}$ or $ts$ is odd.
\end{theorem}      

\proof
Let $\nu \in \mathbb{F}_q^*$ be an element of order $n,$ where $n|k.$ 
Since, $p^t \equiv -1 \pmod{k},$ it follows that $p^t \equiv -1 \pmod{n}$. 
Thus, the equation $p^x \equiv -1 \pmod{n}$ has a positive integer solution $x$. 
Let $t^{\prime}$ be the smallest such solution. 
There exists unique integers $y,r \geq 0$ such that $t = yt^{\prime} + r,$ $r<t^{\prime}.$ Furthermore, 
\[-1 \equiv p^t = p^{yt^{\prime} + r} \equiv (-1)^yp^r \pmod{n}.\]
Since $r<t^{\prime}$, the above equation is only possible if $r = 0.$ Hence, $t^{\prime}|t.$

Now, by Theorem \ref{pure Gauss}, there exists a positive integer $s^{\prime}$ such that $m = 2t^{\prime}s^{\prime},$ so that $2t^{\prime}s^{\prime} = 2ts = 2yt^{\prime}s,$ and hence $s^{\prime} = ys.$ 
Consequently, we have 
\[s \equiv 0 \pmod{2} \implies s^{\prime} \equiv 0 \pmod{2}.\] 
Further, since $ts = t^{\prime}s^{\prime},$ we have 
\[ts \equiv 1 \pmod{2} \implies t^{\prime}s^{\prime} \equiv 1 \pmod{2}.\]

So, it follows from Lemma 4.1  that the conditions guaranteeing that $I_{\beta}(x)|S_2(x)$ are also sufficient to guarantee that $I_{\nu}(x)|S_2(x),$ where $\nu$ is any element of order dividing $k.$ 
Thus, these conditions are sufficient to guarantee that $1 + x + \cdots + x^{k-1}|S_2(x).$ 
And, of course, they are also necessary. The result follows.
$\qed$

We now give some examples to illustrate Theorem \ref{nice thm}. 

\begin{example} Let $p=19$ and let $\mathbf{s}$ be the SLCE sequence of length $19^2 - 1 = 360$ with corresponding polynomial $S_2(x).$ 
Note that $5|20 = 19 + 1.$ Thus, we have $p \equiv 3 \pmod{4}$  and  $s = t = 1$.
Hence, $ts$ is odd. Thus, {\em Theorem \ref{nice thm}} guarantees that $1+x+x^2+x^3+x^4|\text{gcd}(S_2(x),x^{360}+1).$
\end{example}

We  use {\rm Theorem \ref{nice thm}} to interpret some of the numerical results from \cite{K1}. 

\begin{example} 
Let $q=5^2$. 
The authors of \cite{K1} found {\em (}via computer computations{\em )} that $\text{gcd}(S_2(x),x^{q-1}+1) =  (x+1)^4.$ Hence, even though $3|5+1,$ $1+x+x^2 \nmid \text{gcd}(S_2(x),x^{q-1}+1).$ 
Of course, this follows from {\em Theorem \ref{nice thm}} since $p \equiv 1 \pmod{4},$ but $s = 1 \equiv 1 \pmod{2}.$

Let $q = 3^4$. Note that $5|3^2 + 1$ but $5\nmid 3+1$. 
So, $p \equiv 3 \pmod{4}$, $t = 2$ and $s = 1$. Hence, $s \equiv 1 \pmod{2}$ and $ts$ is even, 
so that $1 + x + x^2 + x^3 + x^4 \nmid \text{gcd}(S_2(x),x^{q-1}+1).$ 
This agrees with the calculations in \cite{K1}, where it was found that $\text{gcd}(S_2(x),x^{q-1}+1) = (x+1)^{10}.$

Let $q = 5^4.$ Note that $13|5^2 + 1$  but $13\nmid 5+1$. 
So,  $t = 2,$ $s = 1,$ and $p \equiv 1 \pmod{4}.$ 
Since $s \not\equiv 0 \pmod{2},$ {\em Theorem \ref{nice thm}} guarantees that $1 + x + \cdot \cdot \cdot + x^{13} \nmid S_2(x).$ 
This agrees with the calculations in \cite{K1}, where it was shown that $\text{gcd}(S_2(x),x^{q-1}+1) =  (x+1)^{12}(x^2+x+1)^{10}.$

Let $q = 7^4.$ Note that $5|7^2 + 1.$ So,  $t = 2,$ $s = 1,$ and $p \equiv 3 \pmod{4}.$ 
By {\em Theorem \ref{nice thm}}, since $s \not\equiv 0 \pmod{2}$ and $ts$ is even, $1 + x + x^2 + x^3 + x^4 \nmid S_2(x).$ 
This agrees with the calculations in \cite{K1}, where it was found that $\text{gcd}(S_2(x),x^{q-1}+1) = (x+1)^{22}(x^2+x+1)^{18}(x^4 + x + 1)^2(x^4 + x^3 + 1)^2.$

Let $q = 3^6.$ Note that $7|3^3 + 1.$ So,  $t = 3,$ $s = 1,$ and $p \equiv 3 \pmod{4}.$ 
Thus, $ts$ is odd, and so {\em Theorem \ref{nice thm}} guarantees that $1 + x + \cdots + x^6 \mid S_2(x).$ 
This agrees with the calculations in \cite{K1}, where it was shown that 
\begin{eqnarray*}
&&\hspace{-10mm}\text{gcd}(S_2(x),x^{q-1}+1) \\
&&= (x+1)^2(x^3+x+1)^4(x^3+x^2+1)^4(x^{12} + x^{11} + \cdots + x + 1)^2 \\
&&= (x+1)^2 (1 + x + \cdots + x^6)^4(x^{12} + \cdots + x + 1)^2.
\end{eqnarray*}

Let $q = 5^6.$ Now $3|5+1.$ In this case, $t = 1,$ $s = 3,$ and $p \equiv 1 \pmod{4}.$ 
So, by {\em Theorem \ref{nice thm}}, $1 + x + x^2 \nmid S_2(x).$ 
Also, $3^2|5^3 + 1.$ Here, $t = 3$ and $s = 1.$ 
So, by {\em Theorem \ref{nice thm}}, $1 + x + \cdots + x^8 \nmid S_2(x).$ 
Finally, $7|5^3+1.$ Here, $t = 3,$ and $s = 1.$ So, by {\em Theorem \ref{nice thm}}, 
$1 + x + \cdots + x^6 \nmid S_2(x).$ 
This agrees with the calculations in \cite{K1}, where it was found that 
\begin{eqnarray*}
&&\hspace{-10mm}\text{gcd}(S_2(x),x^{q-1}+1)
= (x^5 + x^3 + x^2 + x + 1)^4(x^5 + x^4 + x^3 + x^2 + 1)^4\\
&&\hspace{40mm} \times (x^5 + x^4 + x^3 + x + 1)^4(x^5 + x^4 + x^3 + x^2 + 1)^4.
\end{eqnarray*}

Let $q = 3^8.$ Now, $5|3^2 + 1.$ Here, $t = 2,$ $s = 2,$ and $p \equiv 3 \pmod{4}.$ 
Hence, since $s \equiv 0 \pmod{2},$ {\em Theorem \ref{nice thm}} guarantees that $1 + x + x^2 + x^3 + x^4|S_2(x).$ 
Also, $41|3^4 + 1.$ Here, $t = 4,$ and $s = 1.$ 
Hence, since $s \not\equiv 0 \pmod{2}$ and since $ts$ is even, 
{\em Theorem \ref{nice thm}} guarantees that $1 + x + \cdots + x^{40} \nmid S_2(x).$ This agrees with the calculations in \cite{K1}, 
where it was shown that 
$$\text{gcd}(S_2(x),x^{q-1}+1) = (x+1)^{26}(x^4 + x^3 + x^2 + x + 1)^{18}.$$ 
\end{example}

We now apply the evaluations of the Jacobi sums of index $2$ given in Corollary~\ref{Jacobi Evaluations} to deduce new divisibility conditions.

\begin{lemma} 
Let $k = \ell^{r}$, where  $\ell $  is a prime congruent to $7 \pmod{8}$ and $r$ is a positive integer.  
We suppose that $[\mathbb{Z}/k\mathbb{Z}:\langle p \rangle ] = 2$ and $m = \phi(k)s/2$, where $s$ is a positive integer.   
Let $e = \phi(k)/2,$ so that $m = es.$ Let $a$ and $b$ be determined as in  Theorem {\rm 2.10} (Langevin's result).

If $p \equiv 1 \pmod{4}$, then \[I_{\beta}(x)|S_2(x) \iff (-1)^{s-1-(p-1)s/4}\left(\frac{a+b}{2}\right)^s \equiv 3 \pmod{4}.\] 

If $p \equiv 3 \pmod{4}$, then \[I_{\beta}(x)|S_2(x) \iff (-1)^{s-1-rs + es +(1-h)s/2}\left(\frac{a+b}{2}\right)^s \equiv 3 \pmod{4}.\]
\end{lemma}

\proof 
Since $\ell \equiv 3 \pmod{4},$ Theorem \ref{quadratic subfield} implies that $K(\chi) \in \mathbb{Q}(\sqrt{-\ell}).$ 
Since $\mathcal{P}$ is a prime ideal lying over $2,$ $\mathcal{P}\cap \mathbb{Q}(\sqrt{-\ell})$ is a prime ideal of $\mathbb{Q}(\sqrt{-\ell})$ lying over $2$ (and conversely, for every prime ideal $\mathcal{P}^{\prime}$ of $\mathbb{Q}(\sqrt{-\ell})$ lying above $2,$ there is a prime ideal $\mathcal{Q}$ of $\mathbb{Q}(\zeta_k)$ lying above $2$ for which $\mathcal{Q}\cap \mathbb{Q}(\sqrt{-\ell}) = \mathcal{P}^{\prime}$). Also, note that the procedure we have outlined in this paper allows us free choice as to which prime ideal of $\mathbb{Q}(\zeta_k)$ lying above $2$ we choose as $\mathcal{P}.$ Finally, recall that an explicit description of the prime ideals lying above $2$ in $\mathbb{Q}(\sqrt{-\ell})$ is given in Theorem \ref{prime ideals}. Without loss of generality, let us choose $\mathcal{P}$ so that 
$$\mathcal{P} \cap \mathbb{Q}(\sqrt{-\ell}) = \langle 2, \frac{-1 + \sqrt{-\ell}}{2}\rangle.$$

In what follows, we will use the fact, mentioned above under Theorem \ref{Langevin}, that $a \equiv b \pmod{2}$ (where $a$ and $b$ are determined as in Theorem \ref{Langevin}) as well as the simple facts that 
$$\frac{1}{2}(K(\chi) + 1) \equiv 0 \pmod{\mathcal{P}} \iff K(\chi) + 1 \equiv 0 \pmod{2\mathcal{P}}$$ 
and that the squares mod $8$ are congruent to either $0,$ $1,$ or $4.$

Since $p \equiv 1,3 \pmod{4},$ it follows that $p^h \equiv 1,3 \pmod{4}.$ Hence, $4p^h \equiv 4 \pmod{8}.$ If $a$ and $b$ are both odd, then $a^2, b^2 \equiv 1 \pmod{8}.$ So, if we assume that this is the case, then by Theorem \ref{Langevin}, 
\[4 \equiv 4p^h = a^2 + \ell b^2 \equiv 1 + 7 \cdot 1 \equiv 0 \pmod{8},\]
which is clearly impossible. Consequently, $a, b \equiv 0 \pmod{2}.$

Case $1$: $p \equiv 1 \pmod{4}.$ By Corollary \ref{Jacobi Evaluations}, we have 
\begin{eqnarray*}
&K(\chi) + 1 &= 1 + (-1)^{s-1-(p-1)s/4}p^{(e-h)s/2}\left(\frac{a + b\sqrt{-\ell}}{2}\right)^s  \\
&&= 1 + (-1)^{s-1-(p-1)s/4}p^{(e-h)s/2}\left(\frac{a+b}{2} + b \left(\frac{-1 + \sqrt{-\ell}}{2}\right)\right)^s.
\end{eqnarray*}
Now, since $2\mathcal{P}|\langle 4 \rangle,$ it follows that $p^{(e-h)s/2} \equiv 1 \pmod{2\mathcal{P}}.$ Further, since $b \equiv 0 \pmod{2}$ and since, by Theorem \ref{AI}, $\frac{-1+\sqrt{-\ell}}{2} \in \mathbb{Z}[\sqrt{n}],$ we have that $b\left(\frac{-1 + \sqrt{-\ell}}{2}\right) \equiv 0 \pmod{2\mathcal{P}}.$ Hence, 
\[K(\chi) + 1 \equiv 1 + (-1)^{s-1-(p-1)s/4}\left(\frac{a+b}{2}\right)^s \pmod{2\mathcal{P}}.\]
But $1 + (-1)^{s-1-(p-1)s/4}\left(\frac{a+b}{2}\right)^s \in \mathbb{Z},$ and $2\mathcal{P} \cap \mathbb{Z} = \langle 4 \rangle.$ Consequently, 
\[I_{\beta}(x)|S_2(x) \iff (-1)^{s-1-(p-1)s/4}\left(\frac{a+b}{2}\right)^s \equiv 3 \pmod{4}.\]

Case $2$: $p \equiv 3 \pmod{4}.$ By Corollary \ref{Jacobi Evaluations}, we have 
\begin{eqnarray*}
&K(\chi)  + 1 &= 1 + (-1)^{s-1-rs + (e + 1)s/2}p^{(e-h)s/2}\left(\frac{a + b\sqrt{-\ell}}{2}\right)^s  \\
&&= 1 + (-1)^{s-1-rs + (e + 1)s/2}p^{(e-h)s/2}\left(\frac{a+b}{2} + b \left(\frac{-1 + \sqrt{-\ell}}{2}\right)\right)^s.
\end{eqnarray*}
Now, since $2\mathcal{P}|\langle 4 \rangle,$ it follows that $p^{(e-h)s/2} \equiv (-1)^{(e-h)s/2} \pmod{2\mathcal{P}}.$ Further, since $b \equiv 0 \pmod{2}$ and since, by Theorem \ref{AI}, $\frac{-1+\sqrt{-\ell}}{2} \in \mathbb{Z}[\sqrt{n}],$ we have that $b\left(\frac{-1 + \sqrt{-\ell}}{2}\right) \equiv 0 \pmod{2\mathcal{P}}.$ Hence, 
\[K(\chi) + 1 \equiv 1 + (-1)^{s-1-rs + es +(1-h)s/2}\left(\frac{a+b}{2}\right)^s \pmod{2\mathcal{P}}.\]
But $1 + (-1)^{s-1-rs + es +(1-h)s/2}\left(\frac{a+b}{2}\right)^s \pmod{2\mathcal{P}} \in \mathbb{Z},$ and $2\mathcal{P} \cap \mathbb{Z} = \langle 4 \rangle.$ Consequently, 
\[I_{\beta}(x)|S_2(x) \iff (-1)^{s-1-rs + es +(1-h)s/2}\left(\frac{a+b}{2}\right)^s \equiv 3 \pmod{4}.\qed\]

Let us now focus on the special case in which $r=1,$ so that $k = \ell.$ 

\begin{theorem} \label{lengthy conditions} 
Let   $\ell \equiv 7 \pmod{8}$ be a prime, and let  $k=l$.  
We suppose that $[\mathbb{Z}/k\mathbb{Z}:\langle p \rangle ] = 2$ and $m = \phi(k)s/2$, where $s$ is a positive integer.   
Let $e = \phi(k)/2,$ so that $m = es.$ Let $a$ and $b$ be determined as in  Theorem {\rm 2.10} (Langevin's result). 

If $p \equiv 1 \pmod{4} \text{ and } b \equiv 0 \pmod{4}$, then 
\[1 + x + \cdot \cdot \cdot + x^{\ell-1}|S_2(x) \iff (-1)^{s-1-(p-1)s/4}\left(\frac{a+b}{2}\right)^s \equiv 3 \pmod{4}.\] 

If $p \equiv 3 \pmod{4} \text{ and } b \equiv 0 \pmod{4}$, then 
\[1 + x + \cdot \cdot \cdot + x^{\ell-1}|S_2(x) \iff (-1)^{s-1-rs + es +(1-h)s/2}\left(\frac{a+b}{2}\right)^s \equiv 3 \pmod{4}.\]
\end{theorem}  

\proof 
Note that $1 + x + \cdot \cdot \cdot + x^{\ell-1}$ is the product of the minimal polynomials of the elements of $\mathbb{F}_{2^f}$ of order $\ell.$ So, if we can guarantee that the relevant condition 
from Lemme 4.2 is the same for each element $\beta$ of order $\ell,$ then we can deduce conditions under which $1 + x + \cdot \cdot \cdot + x^{\ell-1}|S_2(x).$

The explicit conditions given in  Theorem 2.10 are sufficient to determine $a$ completely and to determine $b$ up to sign. In order to determine the sign of $b,$ one must use Stickleberger's congruence  \cite[Lemma $3.5$]{F1}. 
However, we cannot guarantee that the sign of $b$ will be same for Gauss/Jacobi sums corresponding to different characters of order $k$  \cite[Section $11.2$]{B2}. 
But, if we assume that $b \equiv 0 \pmod{4},$ then the residue class mod $4$ of $\frac{a+b}{2}$ is unaffected by the sign of $b$. 
$\qed$

We now give an example to illustrate Theorem \ref{lengthy conditions}. 

\begin{example}
Let $\ell = 23 \equiv 7 \pmod{8},$ let $p = 13 \equiv 1 \pmod{4},$ and let $s = 1.$ It is easy to check that $[(\mathbb{Z}/23\mathbb{Z})^*:\langle 13 \rangle] = 2.$ In this case, $m = \phi(23)/2 = 11,$ so that $q = 13^{11}.$ Referring to the class number table on \cite[p. 325]{A5}, 
we see that $h = h(\mathbb{Q}(\sqrt{-23})) = 3.$ Further, $4p^h = 4 \cdot 13^3 = (74)^2 + 23 \cdot (12)^2,$ so that $a = \pm 74$ and $b = \pm 12,$ and since $a \equiv -2p^{\frac{1}{2}(m+h)}\pmod{\ell},$ we have that $a = 74.$ 
By {\em Theorem \ref{lengthy conditions}}, we have 
\[1 + x + \cdot \cdot \cdot + x^{22}|S_2(x) \iff (-1)^{1-1-(13-1)\cdot 1/4}\left(\frac{74 \pm 12}{2}\right) \equiv 3 \pmod{4} \] \[\iff -37 \equiv 3 \pmod{4}.\]
But $-37 \equiv 3 \pmod{4},$ and so $1 + x + \cdot \cdot \cdot + x^{22}|S_2(x).$
\end{example}

We conclude with a few remarks regarding the applicability of Theorem \ref{lengthy conditions}. 
The fastest way to compute the class number of $\mathbb{Q}(\sqrt{-\ell})$ is via an algorithm due to Shanks, 
which requires at most $O(\ell^{1/4 + \epsilon})$ operations, where $\epsilon$ is any positive number; see \cite[Section 5.4]{C2}. 
The class number of $\mathbb{Q}(\sqrt{-\ell})$ can be used to obtain divisibility results whenever $p$ satisfies $[(\mathbb{Z}/\ell \mathbb{Z}):\langle p \rangle] = 2,$ and it follows by Dirichlet's Theorem on primes in an arithmetic progression that there are infinitely many primes $p$ for which this is true. When the class number  $h = 1,$ there exists a probabilistic polynomial time algorithm, known as the modified Cornacchia algorithm, that can be used to find the integers $a$ and $b$ satisfying $4p^h = 4p = a^2 + \ell b^2;$ see \cite[Section 1.5.2]{C2}. 
In the general case, Hardy, Muskat, and Williams have given a deterministic algorithm that finds $a$ and $b$ (up to sign) in at most $O((4p^h)^{1/4}(\text{log}4p^h)^3(\text{log}\text{log}4p^h)(\text{log}\text{log}\text{log}(4p^h)))$ operations \cite{H4}.

\section*{Acknowledgments} 
The research of \c{S}aban Alaca was supported 
by a  Discovery Grant from the Natural Sciences and Engineering Research Council of Canada (RGPIN-2015-05208),  and 
the research of Goldwyn Millar was supported  by an Ontario Graduate Scholarship.

\vspace{2mm}

\noindent
%Centre for Research in Algebra and Number Theory\\
School of Mathematics and Statistics\\
Carleton University\\
Ottawa, Ontario, Canada K1S 5B6

\vspace{2mm}

\noindent
e-mail addresses : \\
salaca@math.carleton.ca\\
goldwynmillar@cmail.carleton.ca


\begin{thebibliography}{99}

\bibitem{A4} S. Akiyama, \emph{On the pure Jacobi sums}, Acta Arithmetica, LXXV.2, 97-104, 1996. 

\bibitem{A5} S. Alaca and K. Williams, \emph{Introductory Algebraic Number Theory}, Cambridge UP, 2004.

\bibitem{A1} H. Aly and W. Meidl, \emph{On the linear complexity and k-error linear complexity over $\mathbb{F}_p$ of the d-ary Sidelnikov sequence}, IEEE Trans. Inform. Th., Vol. 53 \textbf{12}, 4755 - 4761, 2007.

\bibitem{A2} H. Aly and A. Winterhof, \emph{On the k-Error Linear Complexity over $\mathbb{F}_p$ of Legendre and Sidelnikov Sequences}, Des. Codes Cryptogr. Vol. 40 \textbf{3}, 369-374, 2006.

\bibitem{A3} K. T. Arasu, C. Ding, T. Helleseth, V. Kumar, and H. M. Martinsen, \emph{Almost difference sets and their sequences with optimal autocorrelation}, IEEE Trans. Inform. Theory, vol. 47 \textbf{7}, 2934-2943, Nov. 2001.

\bibitem{Br4} B. C. Berndt and R. J. Evans, \emph{Sums of Gauss, Eisenstein, Jacobi, Jacobsthal, and Brewer}, Illinois Journal of Mathematics, Vol. 23 \textbf{3}, 374-437, 1979.

\bibitem{B2} B. C. Berndt, R. J. Evans, and K. S. Williams, \emph{Gauss and Jacobi sums}, A Wiley-Interscience Publication, 1998. 

\bibitem{B3} T. Beth, D. Jungnickel, and H. Lenz, \emph{Design theory}, Vol. 1, 2nd Edition, Cambridge UP, 1999.

\bibitem{B1} N. Brandst\"atter and W. Meidl \emph{On the linear complexity of Sidelnikov sequences over $\mathbb{F}_d$}, Sequences and their applications - SETA 2006, 47 - 60, Lecture Notes in Comput. Sci., 4086, Springer, Berlin, 2006.

\bibitem{B4} N. Brandst\"atter and W. Meidl, \emph{On the linear complexity of Sidelnikov sequences over nonprime fields}, J. Complexity 24 \textbf{5-6}, 648 - 659, 2008.

\bibitem{B5} N. Brandst\"atter and A. Winterhof, \emph{k-error linear complexity over $\mathbb{F}_p$ of subsequences of Sidelnikov sequences of period $(p^r - 1)/3$}, J. Math. Cryptol., Vol. 3 \textbf{3}, 215 - 225, 2009.

\bibitem{C1} J. H. Chung and K. Yang, \emph{Bounds on the linear complexity and the 1-error linear complexity over $\mathbb{F}_p$ of M-ary Sidelnikov sequences}, Sequences and their applications - SETA 2006, 74 - 87, Lecture Notes in Comput. Sci., 4086, Springer, Berlin, 2006.

\bibitem{C2} H. Cohen, \emph{A Course in Computational Algebraic Number Theory}, Springer-Verlag, Berlin, 1993.

\bibitem{E1} R. Evans, H. D. L. Hollmann, C. Krattenthaler, and Q. Xiang, \emph{Gauss Sums, Jacobi Sums, and p-Ranks of Cyclic Difference Sets}, Journal of Combinatorial Theory, Series A, \textbf{87}, 74-119 (1999).

\bibitem{F1} T. Feng and Q. Xiang, \emph{Cyclotomic constructions of skew Hadamard difference sets}, Journal of Combinatorial Theory, Series A \textbf{119}, 245-256, 2012.

\bibitem{G2} M. Z. Garaev, F. Luca, I. E. Shparlinski, and A. Winterhof, \emph{On the Lower Bound of the Linear Complexity over $\mathbb{F}_p$ of Sidelnikov Sequences}, IEEE Trans. Inform. Th., Vol. 52 \textbf{7}, 3299-3304, 2006.

\bibitem{G1} S. Golomb and G. Gong, \emph{Signal design for good correlation: for wireless communication, cryptography, and radar}, Cambridge UP, 2005.

\bibitem{H2} T. Helleseth, S. H. Kim, and J. S. No, \emph{Linear Complexity over $\mathbb{F}_p$ and Trace Representation of Lempel-Cohn-Eastman Sequences}, IEEE Trans. Inform. Th., Vol 49 \textbf{6}, 1548-1552, 2003.

\bibitem{H3} T. Helleseth, M. Maas, J. E. Mathiassen, T. Segers, \emph{Linear Complexity Over $\mathbb{F}_p$ of Sidel'nikov Sequences}, IEEE Trans. Inform. Th., Vol. 50 \textbf{10}, 2468-2472, 2004.

\bibitem{H1} T. Helleseth and K. Yang, \emph{On binary sequences of period $n = p^m - 1$ with optimal autocorrelation}, Proceedings of SETA$01$ (T. Helleseth, P. Kumar, and K. Yang, eds.), 209-217, 2002.

\bibitem{H4} K. Hardy, J. B. Muskat, and K. S. Williams, \emph{A Deterministic Algorithm for Solving $n = fu^2 + gv^2$ in Coprime Integers $u$ and $v$}, Math. Comp. \textbf{91}, Vol. 55, 327-343, 1990. 

\bibitem{I1} K. Ireland and M. Rosen, \emph{A classical introduction to modern number theory}, 2nd Edition, Springer-Verlag, 1990.

\bibitem{K2} Y. S. Kim, J. S. Chung, J. S. No, and H. Chung, \emph{Linear complexity over $\mathbb{F}_p$ of ternary Sidelnikov sequences} Sequences and their applications - SETA 2006, 61 - 73, Lecture Notes in Comput. Sci., 4086, Springer, Berlin, 2006.

\bibitem{K1} G. Kyureghyan and A. Pott, \emph{On the Linear Complexity of the Sidelnikov-Lempel-Cohn-Eastman Sequences}, Designs, Codes, and Cryptography, 29, 149-164, 2003.

\bibitem{L3} P. Langevin, \emph{Calculs de Certaines Sommes de Gauss}, Journal of Number Theory \textbf{63}, 59-64, 1997.

\bibitem{L1} A. Lempel, M. Cohn, and W. L. Eastman, \emph{A class of binary sequences with optimal autocorrelation properties}, IEEE Trans. Inform. Theory, vol IT-23, 38-42, Jan. 1977.

\bibitem{L2} K. H. Leung and B. Schmidt, \emph{The Field Descent Method}, Designs, Codes, and Cryptography, 171-188, 2005.

\bibitem{M1} S. L. Ma, \emph{A Survey of Partial Difference Sets}, Designs, Codes, and Cryptography, 221-261, 1994.

\bibitem{M5} J. MacWilliams and H. B. Mann, \emph{On the p-rank of the design matrix of a difference set}, Inform. Control \textbf{12}, 474-488, 1968.

\bibitem{M2} H. B. Mann, \emph{Introduction to Algebraic Number Theory}, Ohio State Press, Columbus, Ohio, 1955.

\bibitem{M3} W. Meidl and A. Winterhof, \emph{Some Notes on the Linear Complexity of Sidel'nikov-Lempel-Cohn-Eastman Sequences}, Designs, Codes, and Cryptography \textbf{8}, 159-178, 2006.

\bibitem{S2} K. Shiratani and M. Yamada, \emph{On Rationality of Jacobi Sums}, Colloq. Math., Vol. 73 \textbf{2}, 251-260, 1997.

\bibitem{S1} V. M. Sidelnikov, \emph{Some k-valued pseudo-random sequences and nearly equidistant codes}, Probl. Inform. Trans., vol. 5, no. 1, 12-16, 1969.

\bibitem{X1} L. Xia and J. Yang, \emph{Complete Solving of Explicit Evaluation of Gauss Sums in the Index $2$ Case}, Sci China Math., Vol 53 \textbf{9}, 2525-2542, 2010.
\end{thebibliography}
\end{document}